# THE MODIFIED VLASOV EQUATIONS

**Evangelos Chaliasos**

365 Thebes Street

GR-12241 Aegaleo

Athens, Greece

*Abstract*

The Vlasov system of equations for a plasma is given in relativistic form, and using the correct expression for the "Lorentz" force, that is the one guaranteing real self-consistency.

## 1. Introduction

In a previous paper [1] we gave the really self-consistent system of equations for a two-component plasma. We used the two-fluid approximation and we treated the plasma in full generality (relativistically).

Alternatively, we can use for the plasma the microscopic description by distribution functions. In this way, we have to modify appropriately the Vlasov system of equations [2], first by writting it in relativistic form, and second (the most significant) by the introduction of the correct formula [1] for the "Lorentz" force. This task is undertaken in the present short paper.

In the following, the numbers (of the text) inside parentheses refer to [1].

## 2. The equations

The force needed to be introduced in the collisionless Boltzmann equations (the Vlasov equations for the distribution functions) must not be given by (35) but rather by (36), as I mention in my paper [1]. Thus, it is suggested to use the potentials instead of the fields themselves. In this way, and in conventional notation, this force reduces to

$$\vec{f} = -\frac{q}{c}\frac{\partial \vec{A}}{\partial t} - \frac{q}{c}(\vec{v}\cdot grad)\vec{A}. \tag{a}$$

Further, in order to have relativistic equations, it is suggested to use the momenta rather than the velocities, the latter given by

$$\vec{v} = \frac{\vec{p}}{\sqrt{m^2 + p^2/c^2}}. \tag{b}$$

Then the Vlasov equations under consideration become

$$\frac{\partial f_\pm}{\partial t} + \frac{\vec{p}}{\sqrt{m_\pm^2 + p^2/c^2}} \cdot \frac{\partial f_\pm}{\partial \vec{r}} - \frac{q_\pm}{c}\left[\frac{\partial \vec{A}}{\partial t} + \left(\frac{\vec{p}}{\sqrt{m_\pm^2 + p^2/c^2}} \cdot grad\right)\vec{A}\right] \cdot \frac{\partial f_\pm}{\partial \vec{p}} = 0. \tag{c}$$

These equations must of course be complemented by the Maxwell equations in terms of the potentials in a particular gauge. These are just (31), in the gauge (30). Thus we are left in conventional form with

$$\frac{1}{c^2}\frac{\partial^2 \phi}{\partial t^2} - \nabla^2 \phi = 4\pi\rho \tag{d1}$$

$$\frac{1}{c^2}\frac{\partial^2 \vec{A}}{\partial t^2} - \nabla^2 \vec{A} = \frac{4\pi}{c}\vec{j} \tag{d2}$$

(Maxwell equations), and

(Lorentz gauge).

$$\frac{\partial \phi}{c\partial t} + div\vec{A} = 0 \tag{e}$$

Finally, the (total) current in (d) has also to be taken from the equations

$$\rho = \int (q_+ f_+ + q_- f_-)d^3 p, \tag{f}$$

$$\vec{j} = \int \left(\frac{q_+ f_+}{\sqrt{m_+^2 + p^2/c^2}} + \frac{q_- f_-}{\sqrt{m_-^2 + p^2/c^2}}\right)\vec{p}d^3 p \tag{g}$$

(these are equations (27.11) of [2] in relativistic version).

I mention that we must also use the continuity equation for one of the two species (we assume that the material is fully ionized and recombinations do not occur), written ( for the electrons, for example) as

$$\frac{\partial}{c\partial t}\int f_- d^3 p + div \int \frac{f_-}{\sqrt{m_-^2 + p^2/c^2}} \vec{p} d^3 p = 0. \tag{h}$$

Thus we now have as the vlasov system of equations the following ones: 2(c) + 4(d) +1(e) + 4(f&g) + 1(h) = 12. The unknowns are: $f_\pm$ (2) + ($\varphi$, **A**) (4) + (c$\rho$, **j**) (4) = 10. That is we have a system of 12 coupled equations for 10 unknowns.

These equations are the really self-consistent system of equations describing the plasma, which have to be taken instead of (27.9)-(27.11) of [2] (pp. 117-118). The novelty of this modified system of Vlasov equations lies basically in the introduction of the correct "Lorentz" force (a) in the two collisionless Boltzmann equations (c).

### 3. Comments

We gave the equations in relativistic form for full generality. If one wants to simplify them by not taking into account relativistic effects, he has to simply omit the term $p^2/c^2$ under the square roots appearing.